\UseRawInputEncoding
\documentclass[superscriptaddress,floats,amsmath,amssymb,prl,aps,twocolumn]{revtex4}

\usepackage{graphics,amsmath,graphicx}
\usepackage{txfonts}
\usepackage[colorlinks=true,citecolor=blue,linkcolor=blue,urlcolor=blue]{hyperref}

\begin{document}


\title{A quantum theory of the nearly frozen charge glass}

\author{S. Fratini$^*$} 
\affiliation{Institut N\'eel, CNRS \& Univ. Grenoble Alpes, 38042 Grenoble, France}

\author{K. Driscoll}
\affiliation{Institut N\'eel, CNRS \& Univ. Grenoble Alpes, 38042 Grenoble, France}

\author{S. Ciuchi}
\affiliation{Institute for Complex Systems, University of L'Aquila, L'Aquila, Italy}

\author{A. Ralko}
\affiliation{Institut N\'eel, CNRS \& Univ. Grenoble Alpes, 38042 Grenoble, France}

\begin{abstract} 
We study long-range interacting electrons on the triangular lattice using mixed quantum/classical simulations going beyond the usual classical descriptions of the lattice Coulomb fluid. Our results in the strong interaction limit indicate that the emergence and proliferation of quantum defects governs the low-temperature dynamics of this strongly frustrated system, in a way that crucially depends on the degree of anisotropy of the electronic structure.
The present theoretical findings explain the phenomenology observed in the $\theta$-ET$_2$X charge ordering materials as they fall out of equilibrium. The approach devised here can be easily generalized to address other  systems where charge frustration is lifted by quantum fluctuations. 
\end{abstract}


\maketitle

\paragraph{Introduction.} 
The organic metals of the $\theta$-ET$_2$X family host a puzzling charge glass state \cite{Nad07,KagawaNphys13} that still lacks a proper microscopic description.
These materials are composed of layers of BEDT-TTF (ET) organic molecules whose ordered arrangement approaches a triangular lattice, with slight deviations that can be tuned by an accurate choice of the cation X. At
the concentration of one hole (three electrons) per two molecules determined by stoechiometry, the triangular geometry is known to frustrate possible electronic orders that would result from the strong Coulomb repulsion between the electrons.  \cite{Hotta06}
The observed charge glass is particular in that it emerges in compounds that are apparently devoid of structural disorder: this implies the existence of some form of self-generated randomness originating from the dynamical arrangement of the electrons themselves. 
It has been proposed that the randomness underlying the glassy behavior could 
arise from 
the competition between the many metastable states that emerge when long-range order is frustrated \cite{Andreev79,efros_coulomb_1992,KagawaNphys13,Mahmoudian15}.

Experimentally, the extensive role 
of metastable configurations was suggested by 
the observation of a progressive freezing of electronic dynamics upon lowering the temperature, accompanied by aging phenomena, as well as the presence of diffuse signals in the X-ray spectra indicative of a continuum of competing (short-range) ordered structures \cite{KagawaNphys13}. 
Many experiments 
have explored in particular the dynamics of the approach to glassiness by making use of thermal quenches in order to avoid electronic crystallization  \cite{Sato14,SatoScience17}. It is now understood that the  velocity of the quench required to access the charge glass varies from material to material, and  correlates to the degree of geometrical frustration set by the molecular lattice. The transport properties are also affected by the amount of frustration, and switching between resistivity values differing by several orders of magnitude can be observed when the system transitions from an electronically ordered to a metastable glassy state \cite{KagawaNphys13,SatoNMat19}. 

The underlying idea that charge frustration could be at the origin of the observed behavior was given a microscopic basis soon after the discovery \cite{Mahmoudian15}, demonstrating the crucial role of \emph{long-range} interactions between the holes. It was  shown that the combination of long-range Coulomb interactions and the triangular geometry of the molecular lattice at one quarter filling leads to the emergence of very many competing metastable states with amorphous “stripe-glass” spatial structures and coexisting short- and long-range order. 
The dynamics of the resulting frustrated phase were explored through classical Monte Carlo (MC) simulations \cite{Mahmoudian15}, which showed remarkably slow viscous dynamics typical of strong glass formers as well as aging phenomena characteristic of supercooled liquids on shorter timescales. This phenomenology, which strikingly resembles the experimental observations, is instead totally absent for short-range interactions. 

If one thing is puzzling about 
the agreement between theory and experiment is that it is based on purely classical simulations. 
The bandwidths in the 
$\theta$-ET$_2$X class range from few tenths to half an eV \cite{Mori}, and it can be argued that the quantum fluctuations associated with fast inter-molecular electron transfer should be able to  melt the glassy state. Even more importantly, it is hard to explain on the basis of purely classical considerations the extreme sensitivity of the glass-forming proneness to only slight modifications of the structure \cite{Sato14,SatoNMat19}.
Even large changes in the lattice anisotropy as obtained by chemical or physical pressure lead to anisotropies in the intermolecular nearest-neighbor repulsion that do not exceed 15$\%$, while the dynamical timescales at play, viz. the speed of quenches needed to drive the system out of equilibrium, vary by orders of magnitude from compound to compound.


Here we show that quantum fluctuations do not immediately melt the glass, provided that intermolecular transfer integrals are modest and 
the Coulomb energy is the dominant term. This corresponds precisely to the strongly interacting regime that applies to the $\theta$-ET$_2$X organic metals.
Promoted by quantum fluctuations, quantum defects emerge and proliferate,  governing the physics of the electronic system  in a way that depends dramatically on the degree of frustration, being controlled by the anisotropy of the band structure. In the maximally frustrated isotropic case,  
quantum defects 
actually favor the stabilization of disordered configurations, by deepening  metastable local minima in the potential energy landscape. For anisotropic band structures, instead, the directionality of inter-molecular overlaps favors the alignment of the defects, stabilizing stripe ordered configurations.
Because the electronic structure (i.e. the intermolecular transfer integrals) rather than the anisotropy of interactions is the microscopic parameter that is mostly sensitive to slight structural changes in organic metals,
the physics of quantum defects proposed here appears as a strong candidate to explain the puzzling phenomenology observed in the compounds of the $\theta$-ET$_2$X class.

\paragraph{Model and method.} 
\begin{figure*}[htpb]
\includegraphics[width=2\columnwidth]{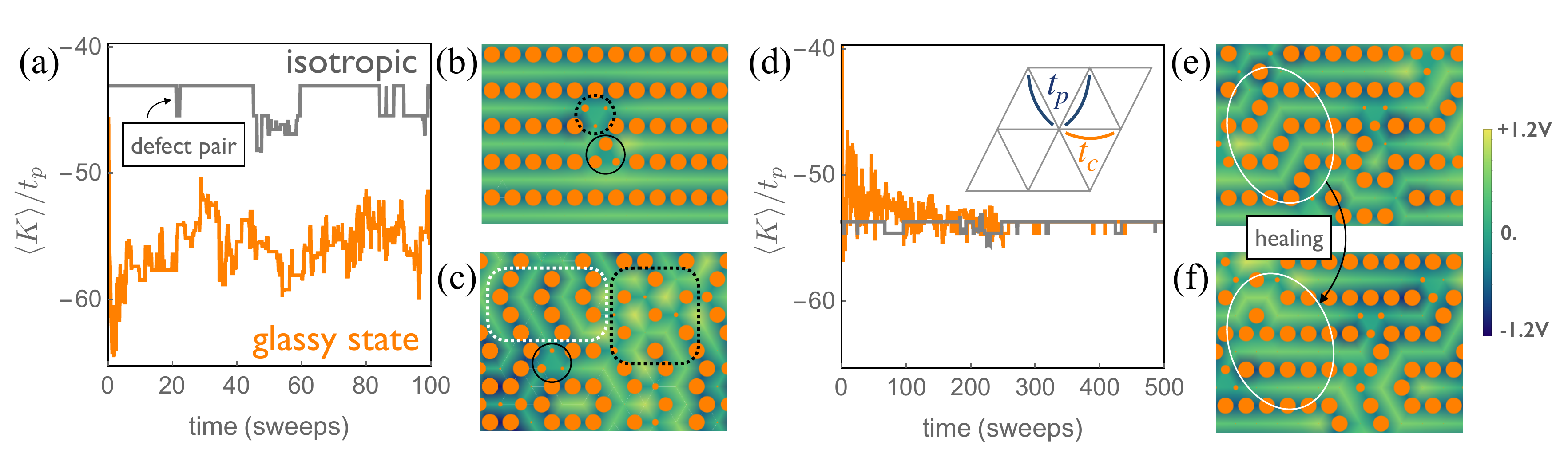}
\caption{{\bf Kinetic evolution and emergence of defects}
(a) Thermalized kinetic energy as a function of MC time for a single walker initially prepared in the perfect horizontal stripe arrangement (gray) or in a maximally disordered configuration (orange). Simulations are performed on a regular $L\times L$ lattice with $L=12$ in the strong coupling regime $t_p=0.09 V$, $\eta=-1$, and at a temperature $T=0.03V$, lower than the classical charge ordering temperature $T_c = 0.038V$ \cite{Mahmoudian15}. Time is measured in sweeps, where one sweep $=L^2$ update attempts. (b,c)  Typical charge configurations: The radius of the orange disks is proportional to the local hole density, the background map is the collective electrostatic potential. (d-f) Kinetic energy evolution and typical configuration for an anisotropic band structure, $t_p=0.12 V$, $\eta=-0.1$. The two configurations shown correspond to respectively $t=95$ and $t=100$, illustrating the process of stripe order "healing" caused by the anisotropic electronic structure.}
 \label{fig:traces}
\end{figure*} 
We study the following Hamiltonian:
\begin{equation}
    H=   \sum_{\langle ij\rangle} t_{ij} \ c^\dagger_i c_j +h.c. +\sum_{ij} V_{ij} \hat{n}_i \hat{n}_j
    \label{eq:Hamiltonian}
\end{equation}
where the first term, hereafter denoted the kinetic term $K$, describes the hopping of  fermions on a lattice,  $\hat{n}_i=c^\dagger_i c_i$ is the local density operator and $V_{ij}=V/|R_{ij}|$ the unscreened Coulomb potential between electrons on molecular sites $R_i$ and $R_j$, that is treated via standard Ewald summations. We consider a triangular lattice  as appropriate for the $\theta$-ET$_2$X class, with a concentration of one hole per two sites and transfer integrals $t_{ij}=t_p,t_c$
respectively in the diagonal and horizontal directions (see Fig. \ref{fig:traces}), neglecting the spin degree of freedom.
We consider values of $t_p/V \ll 1$ and vary independently the anisotropy ratio $\eta = t_c/t_p$ which, for the materials studied by Kanoda and coworkers, ranges in the interval $-1 \lesssim \eta \lesssim 1$ \cite{McKenzie}. The expression Eq. (\ref{eq:Hamiltonian}) has an explicit positive sign for $t_{ij}$ as appropriate for holes.


The thermal behavior of the model Eq. (\ref{eq:Hamiltonian}) was studied in Ref. \cite{Mahmoudian15} through classical MC simulations, highlighting 
a marked tendency to glassy behavior at low temperatures. 
Here we want to address the effects of quantum fluctuations when $t_{ij}\neq 0$.
To do so, we perform a strong coupling perturbation expansion, treating the  kinetic term in Eq. (\ref{eq:Hamiltonian}) to lowest order in $t_p/V$. This methodology was successfully applied to study the quantum melting of the charge ordered state in the presence of long-range Coulomb interactions, both in one \cite{kinks-Baeriswyl,kinks-Rastelli} and two space dimensions \cite{Fratini09,Driscoll}. There it was shown to correctly capture the proliferation of defects involved in the quantum melting mechanism at $T=0$, 
providing results in quantitative agreement with full exact diagonalization.

We  generalize the strong coupling expansion described above to finite temperatures, by coupling it to classical MC simulations to account for the many different charge configurations that are thermally accessible beyond the classical ground state. We use local nearest-neighbor updates to mimic the hopping of electrons between molecules \cite{Kolton}.
At each step in the MC evolution we then solve the one-body electron problem in the electrostatic potential determined by the collective distribution of classical occupations $\{n_i\}=0,1$, hence allowing the hole density to spread away from their classical positions on the molecular sites.
This step is equivalent to the one that was used in the ordered case in Refs. \cite{kinks-Baeriswyl,kinks-Rastelli,Fratini09} (where however only the configuration of minimal energy was considered), giving access to several physical observables such as the spatial distibution of the charge,
the one-particle spectral function and the optical conductivity. Second, we include in the MC engine the kinetic energy gain enabled by the quantum spreading of the charge.
 This is achieved by replacing the classical statistical weight $\exp[-\beta E_{\{n_i\}} ]$ in the Metropolis-Hastings sampling by the quantum corrected weight  $ \exp[-\beta (E_{\{n_i\}} + \langle K \rangle_{\{n_i\}})] $, where $\beta=1/k_BT$, $E_{\{n_i\}}$ is the  energy of the classical configuration and $\langle K \rangle_{\{n_i\}}$ is the thermalized  quantum expectation value of the kinetic term in that same configuration. Mathematically, this rigorously corresponds to the cumulant resummation of the perturbation expansion to lowest order in $K$ \cite{cumulant}. Physically, this means that at each step we work 
within the restricted statistical ensemble associated to a given metastable minimum in the classical configuration space, letting the electrons fully thermalize within this minimum.

The method devised here can be seen as a practical implementation of early ideas of Andreev and Kosevich \cite{Andreev79} and  Efros \cite{efros_coulomb_1992}, 
who hypothesized a clear separation of time-scales 
between fast individual and slow collective motions, the latter being severely slowed down by the existence of many-body interactions at all distances.
This "two fluid" behavior of the frustrated Coulomb system 
has been recently confirmed by a fully quantum study of long-range interacting electrons on the triangular lattice \cite{Driscoll}.

\paragraph{Emergence of quantum defects.} 
Fig. \ref{fig:traces} shows the evolution of the kinetic energy $\langle K \rangle$ together with typical charge configurations and potential energy landscapes explored during the MC simulations in representative cases. In panel (a), the gray line is the evolution of  $\langle K \rangle$ for a system with an isotropic band structure, starting the simulation from a perfectly ordered horizontal stripe configuration. This trace shows abrupt quantized jumps from a constant baseline, signaling the creation and annihilation of defect pairs with energy of order $t_p$.
Their shape is shown in panel (b): Classically they correspond to moving a particle from a site on a charge rich stripe to a neighboring empty stripe. This creates  an electrical dipole  constituted of two oppositely charged defects (full and dashed circle; see \cite{Fratini09} for the analogous defect pairs on the square lattice). Specific to the triangular lattice, this local fluctuation defines a region of three adjacent sites (full circle), all of them having three occupied neighbors and therefore an almost degenerate electrostatic potential $\phi_i=\sum_j V_{ij} n_j$ (up to next nearest neighbor corrections).
Quantum-mechanically, a metallic droplet is formed through hybridization on these three sites, gaining an energy $E_d \propto t_p$.

To mimic a rapid quench starting from high temperature, 
the orange line in Fig. \ref{fig:traces}(a) reports the evolution of $\langle K \rangle$ starting from a random configuration. After a sharp initial drop, the kinetic energy reaches a stationary regime exhibiting large fluctuations of several $t_p$ around an average value that is much lower than that of the ordered stripe configuration: Both features are indicative of the presence and dynamics of a large number of defects. The corresponding  configuration map shown in panel (c) indeed shows finite-size striped domains of random orientations coexisting with a competing threefold order (respectively white and black dotted), in addition to isolated defects (circle). 
Note that the latter naturally arise at the crossings of stripes of three different orientations, which follows from the same neighbor counting argument presented above. 
The  kinetic energy gain associated with threefold stripe crossings, and more generally with the proliferation of more extended quantum defects, is  what makes the stripe glass arrangement particularly stable. Disordered configurations with random stripe orientations such as the one shown in panel (c) are eventually reached at very long times even when starting from the ordered case (not shown).


Fig. \ref{fig:traces}(d) show traces analogous to the ones shown in  Fig. \ref{fig:traces}(a), but now with a highly anisotropic electronic structure,  $\eta=-0.1$. As in the isotropic case, the evolution starting from the ordered state (gray) shows successive creation and annihilation of defects. More interesting is the evolution after a quench (orange): the kinetic energy here steadily decreases and reaches the same baseline value of the ordered state, indicating 
an unforeseen tendency to ordering. This healing capability can be understood from the microscopic process sketched in panels (e,f). Because of the anisotropy of the transfer integrals, the horizontal stripe direction is now favored over the other directions, because it maximizes the kinetic energy gain through transverse fluctuations with probability $\propto (t_p/V)^2 > (t_c/V)^2$ .
The same defects that favored disordered stripe clusters in the isotropic case now have an opposite effect, creating order from a maximally disordered initial state.

\paragraph{Aging.}
\begin{figure}[thpb]
\includegraphics[width=8cm]{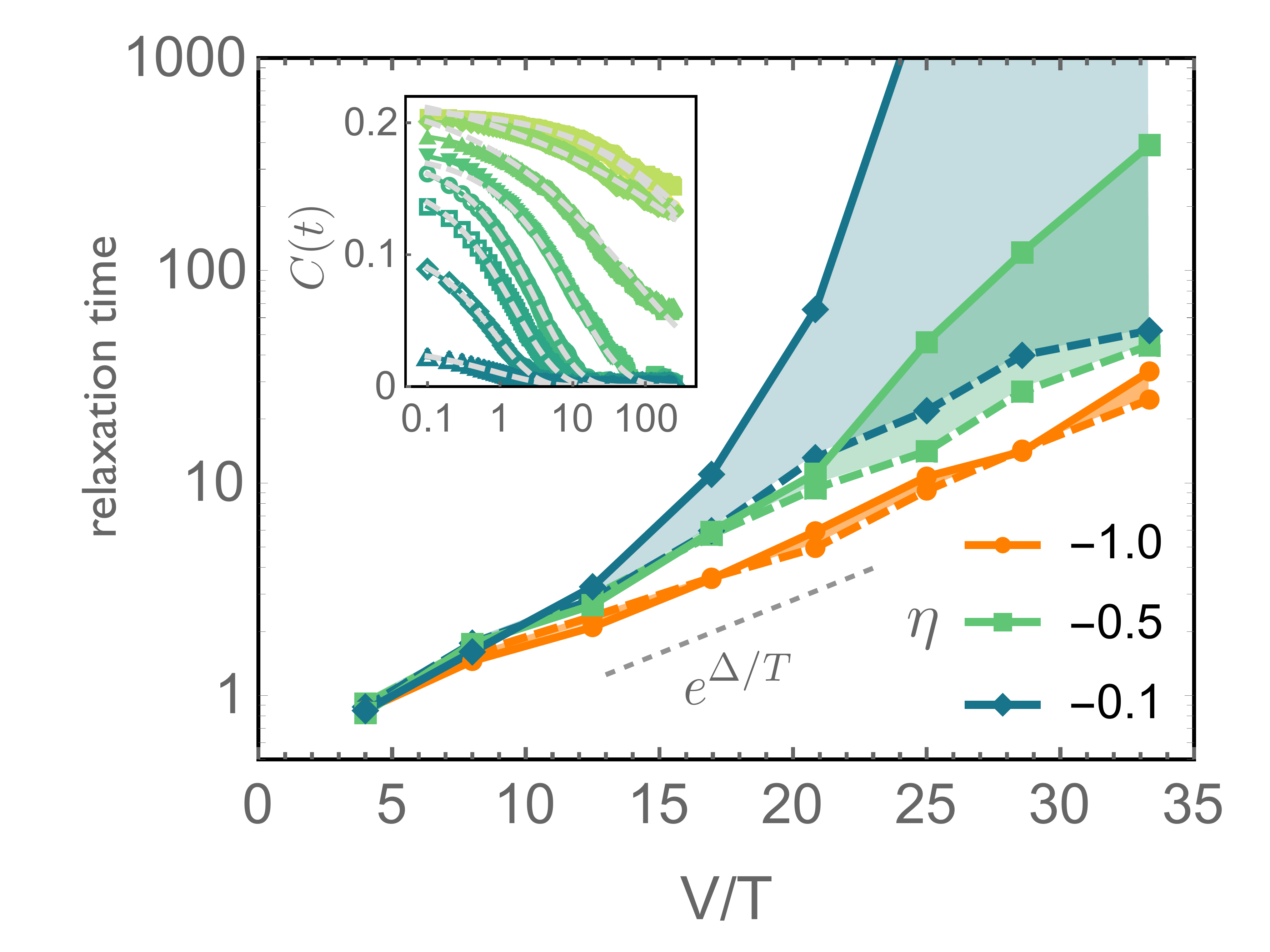}
\caption{ {\bf Relaxation times and aging}. Arrhenius plots of the relaxation times as determined from stretched exponential fits of the local density-density correlation function (inset, temperatures decreasing from bottom to top, fits shown as dashed lines) for different degrees of anisotropy $\eta$.  Dashed and full lines correspond respectively to waiting times $T_W=20,200$.}
 \label{fig:autocorr}
\end{figure} 
The dependence on history highlighted in the previous paragraphs can be assessed quantitatively by studying the local autocorrelation function $C(t,t_W)=\sum_i \langle  n_i(t+t_W)  n_i(t_W) \rangle/N$, providing information on the dynamic relaxation processes \cite{Kolton}. Here $t_W$ is the waiting time  the system is allowed  to relax after initializing it to a random initial configuration, and the average is performed over $\sim$100 independent walkers.

Fig. \ref{fig:autocorr}(a) shows the time dependence of $C$ over 250 sweeps ($t_p=0.12V$, $t_c/t_p=-0.1$), after a waiting time $T_W=200$. Superimposed on the numerical data are stretched exponential fits $C(t,t_W)=C_0 \exp (-(t/\tau)^\alpha)$ with $\alpha<1$. Panel (b) reports the extracted relaxation times for different values of the anisotropy ratio $\eta$ (labels and different colors). For each value of $\eta$, the   lower (dashed) and upper (full) curves correspond to waiting times $T_W=20,200$ sweeps respectively.

For isotropic band structures (orange), the relaxation time grows in an Arrhenius fashion upon cooling, typical of strong glass formers in the supercooled liquid regime (here $\Delta \simeq 0.11V$,  smaller than the value $\Delta \simeq 0.2V$ found in the classical case \cite{Mahmoudian15}). It does so independently on the waiting time, indicating that  equilibration occurs at times faster than $T_W=20$ in the whole  temperature window studied. The situation changes drastically upon introducing electronic anisotropy (green and blue). Here a dynamical crossover temperature emerges (panel (c), orange symbols), below which the system falls out of equilibrium:
the relaxation time depends strongly on the waiting time, growing faster than exponential for long waiting times. This corresponds to the onset of the stripe order healing process described in the preceding paragraphs:  the relaxation time grows as charge fluctuations become less and less frequent upon dynamically approaching the ordered state (correspondingly, the stretched exponential fits become less and less accurate). 

\paragraph{Characterization of the electronic system and resistivity switching.} 
 \begin{figure}[htpb]
 \includegraphics[width=0.95\columnwidth]{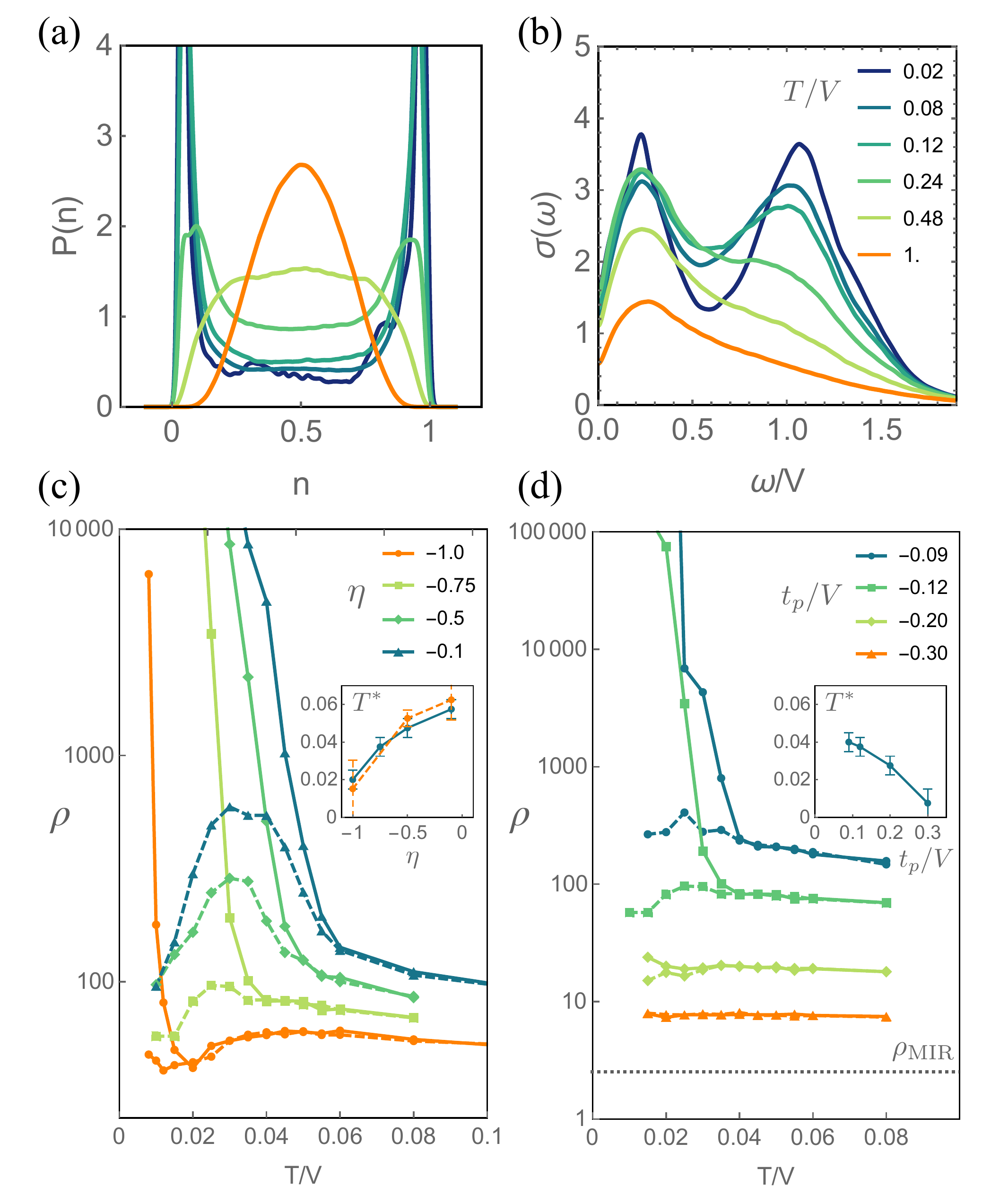}
\caption{ {\bf Electronic properties and switching.}
(a) Distribution of local densities and (b) optical conductivity spectra at different $T/V$ for $\eta=-1$ and $t_p/V=0.12$. (c) Resistivity vs temperature obtained after a waiting time $T_W=200$  starting from a quench (dashed) and an ordered initial state (full),   for different  $\eta$ for $t_p/V=0.12$. (d) resistivity  for $\eta=-0.75$ and different $t_p/V$. Units of $\rho$ are $c \hbar/e^2\simeq 400 \mu \Omega$cm, with $c\simeq 10\AA$ the interlayer distance. The horizontal arrow marks the Mott-Ioffe-Regel limit, $\rho_{\mathrm{MIR}}\simeq 1$ m$\Omega$cm. The insets show the dynamical temperature $T^*$ where the resistivity (blue) and the density correlations (orange dashed) become history dependent.
 }
 \label{fig:transport}
\end{figure} 
Fig. \ref{fig:transport}(a) shows the distribution of local hole densities on the molecular sites, calculated for a waiting time $T_W=200$ after a quench in the isotropic case $\eta=-1$  ($t_p=0.12V$, $T/V$ labels as per panel (b)). At high temperatures the distribution is bell shaped and centered around the average hole concentration $n=1/2$, indicating a normal fluid without charge order. Upon cooling, the distribution progressively becomes bimodal, with sharp peaks at $n=\delta,1-\delta$, with $\delta \propto (t_p/V)^2$ as predicted by second order perturbation theory. The distribution is seen to saturate at very low temperatures due to the persistence of quantum defects, as the system  never reaches the perfect stripe ordered state. This saturation occurs at a temperature   $T \sim t_p$, and is signaled by a residual probability of finding local density values in the whole interval $0<n<1$, in agreement with recent experimental findings \cite{Kanoda-crossover}.

Fig. \ref{fig:transport}(b) shows the  optical conductivity calculated for the same microscopic parameters, by evaluating the Kubo-Greenwood formula for holes in the random electrostatic potential determined by the instantaneous distribution of classical charges.
A Lorentzian broadening $\gamma$ has been added to the spectra to mimic the timescale of the collective charge motions that are at the origin of the random potential, as described in Refs. \cite{Fratini16,scipost}. We take the value $\gamma=0.2 t_p$ as determined from the fully quantum solution of the problem at $T=0$  in \cite{Driscoll}. The spectra  exhibit two separate peaks of different nature: (i) A charge ordering peak at $\omega=V$, corresponding to the local fluctuation of holes in a short-range ordered, striped environment; this peak  is washed out upon heating at $T\sim V$, i.e. when short-range order disappears and (ii) A displaced Drude peak (DDP) at $\omega=2t_p$  that persists at all temperatures, probing the dynamics of holes trapped by the strongly disordered collective electrostatic potential. This peak is one microscopic realization of the disorder-induced DDP that is generally predicted to occur when electronic carriers interact with  slowly fluctuating degrees of freedom \cite{scipost}. The latter are  represented here 
by the collective charge fluctuations of the frustrated Coulomb fluid, as was already suggested in other frustrated organic metals \cite{Pustogow}.
The shape and temperature dependence of the calculated DDP are strongly reminiscent of what is observed in the isotropic material $\theta$-ET$_2$I$_3$ \cite{Takenaka05}.

We finally comment on the transport properties of the model, highlighting a strong dependence on the cooling dynamics as observed in experiments. Figs. \ref{fig:transport}(c,d) show the resistivity curves obtained from the $\omega=0$ limit of the optical conductivity, for (c) varying levels of electronic anisotropy $\eta$ at $t_p=0.12V$  and for (d) varying $t_p$ at fixed $\eta=-0.75$. For each choice of parameters we report results corresponding to a quench (dashed), i.e. instantaneous cooling from a random configuration at $T=\infty$ down to the desired temperature,  and those obtained by starting the simulation from the horizontal stripes that minimize the electrostatic energy (full). The latter prescription favors charge ordering, and it is therefore representative of experiments done at slow cooling rates.

In all cases, the resistivity curves show switching between a low resistance state (quench) and a high resistance state (order) at a temperature $T^*$ that depends crucially on the microscopic parameters $\eta$ and $t_p$, cf. insets in panels (c,d).  Confirming the findings of the preceding paragraphs,  isotropic electronic structures  and large transfer integrals (i.e., large quantum fluctuations) both favor the emergence of quantum defects, leading to a reduction of  $T^*$ and to  an overall decrease of the absolute value of the resistivity.
Almost T-independent resistivity curves are obtained at the approach of the Mott-Ioffe-Regel limit in the most frustrated case. The overall features shown in  Fig. \ref{fig:transport}(c,d) are in qualitative agreement with the experimental measurements in the $\theta$-ET$_2$X series \cite{SatoNMat19}.

\paragraph{Concluding remarks.} 
The results presented here rationalize the extreme sensitivity of the electronic properties of the $\theta$-ET$_2$X salts to chemical strain, 
based on the proliferation and dynamics of quantum defects. The theory highlights the key role played by quantum fluctuations in the competition between disordered metastable and charge ordered states occurring in strongly frustrated electron fluids, which could be of general relevance in other systems of interest.
In this respect, it will be interesting to apply the method devised here to address the "hidden" metastable phase that can be accessed dynamically in 1T-TaS$_2$ close to Wigner crystal ordering  \cite{mihailovic-science}.
Similar to the organic compounds studied here, this system 
also hosts strongly frustrated interacting electrons on a triangular lattice,
and recent STM experiments \cite{ravnik-billiards} have shown interference patterns 
indicating the predominance of quantum defect dynamics beyond the simple classical picture.

\vfill
\pagebreak


\bibliographystyle{aipauth4-1}


$^*$ simone.fratini@neel.cnrs.fr

\end{document}